\documentclass[prc,superscriptaddress]{revtex4}
\usepackage{amssymb,color,graphicx,bm,amsmath}

\definecolor{red}{rgb}{0.8,0,0}
\definecolor{RED}{rgb}{0.8,0,0}
\definecolor{violet}{rgb}{0.4,0,0.4}
\definecolor{green}{rgb}{0,0.5,0.0}
\definecolor{GREEN}{rgb}{0,0.5,0.0}
\definecolor{navy}{rgb}{0.0,0.0,0.6}
\definecolor{orange}{rgb}{0.8,0.2,0.0}
\definecolor{blue}{rgb}{0.3,0.0,0.8}

\begin{document}

\title{Establishing a relation between mass and spin of stellar mass black holes}

\author{Indrani Banerjee, Banibrata Mukhopadhyay$^*$\\
Department of Physics, Indian Institute of Science, 
Bangalore 560012, India\\ indrani@physics.iisc.ernet.in , bm@physics.iisc.ernet.in\\
$^*$Corresponding author
}

\begin{abstract}
Stellar mass black holes (SMBHs), forming by the core collapse of very massive, rapidly rotating stars,
are expected to exhibit a high density accretion disk around them developed from the spinning mantle of 
the collapsing star. A wide class of such disks, due to their high density and temperature, are effective 
emitters of neutrinos
and hence called neutrino cooled disks. Tracking the physics relating the observed (neutrino) luminosity 
to the mass, spin of black holes (BHs) and the accretion rate ($\dot{M}$) of such disks, here we establish 
a correlation between the spin and 
mass of SMBHs at their formation stage.
Our work shows that spinning BHs are more massive 
than non-spinning BHs for a given $\dot{M}$. However, slowly spinning BHs can turn out to 
be more massive than spinning BHs if $\dot{M}$ at their formation stage was higher compared to faster 
spinning BHs.

\end{abstract}

\maketitle


\paragraph*{Introduction.$-$} 

The formation history and exact value of spin for observed black holes (BHs) is still a mystery.
It is believed that the stellar mass black holes (SMBHs) result from the core collapse of massive 
stars leading to a supernova explosion and/or a gamma-ray burst (GRB) \cite{Celloti99}. 
Since the progenitor of such BHs is a 
rapidly rotating massive star, it is often speculated that these BHs 
will be endowed with an intrinsic rotation \cite{MW99}. Moreover, the rotating stellar mantle 
forms an accretion disk around the SMBHs and, as the BH accretes, its mass and spin both evolve.
At the completion of stellar collapse, the mass and spin which the BH attains, 
remain as the fundamental parameters of the newly formed BH.

The Kerr metric \cite{Kerr} describes the geometry of empty spacetime around an uncharged, rotating, axisymmetric 
BH. However, until last decade, there was no
estimate of the spin of the known SMBHs. On the other 
hand, the mass of several SMBHs was determined independently. 
Therefore, the question arises, is it 
possible to correlate the mass and spin of these BHs? If the mass of BHs
is known, which is relatively easy to determine observationally, can one predict their spin? On which other 
parameter(s), if any, do the mass and spin of SMBHs depend? 
These are some of the fundamental questions, we plan to address in the present letter.

\paragraph*{Basic assumptions and governing equations.$-$}

From general relativity, if the radius of the BH is known, then its spin and mass can be 
correlated, which can be understood from the expression for event horizon
\begin{equation}
R_+ = M + (M^2 - a^2)^{1/2},
\end{equation}
where $R_+$ represents the radius of the event horizon, $M$ and $a$ are the mass and specific angular 
momentum 
of the BH respectively. We work in natural units where the gravitational constant $G$ and
the speed of light $c$ are chosen to be unity.
However, we cannot fix $R_+$ uniquely for all BHs. People generally try to estimate $R_+$ from observation
of various BH sources and supplying mass, determined from an independent measurement, obtain their spin. 
Therefore, in order to correlate the spin with mass, event horizon does not serve any generic purpose.

In order to constrain the mass and spin of SMBHs, we plan to consider the properties of the accretion
disk developed around the BH during its formation stage. These disks are associated with a very high 
accretion rate ($\dot{M}$): $0.0001 \lesssim \dot{M}/M_\odot s^{-1} \lesssim 10$, $M_{\odot}$ being the mass of the Sun. 
For $0.01 \lesssim \dot{M}/M_\odot s^{-1} 
\lesssim 10$, the disks are called collapsar~I disks \cite{MW99, MF01}. Here, the core of the 
progenitor directly collapses to a BH with no observed supernova explosion \cite{Fryer99}. 
The temperature exceeds 
$10^{11}$K in the inner region of these disks,
enabling them to cool via neutrino emission. 
Neutrino emission from these disks takes place independent of any GRB being launched from them 
\cite{MacFadyen}.
This emission is completely different from jets driven by neutrino-antineutrino
annihilation. In fact, there can always be a BH formation which is not accompanied by a GRB/jet, however, 
neutrinos can still emanate from the disks surrounding such BHs. All that it requires
for a disk to emit neutrinos is to harbor a temperature significantly above $10^{10}K$. Moreover, it 
was shown that the luminosity of such neutrino emission remains roughly constant \cite{Proga}, lying
in a narrow range
$\sim 10^{52}-10^{53} \rm ~ergs ~ s^{-1}$, once the accretion disk is formed and steady state is achieved.
Similar orders of neutrino luminosity from collapsar~I disks were also estimated by \cite{MW99}. 
Once neutrino cooling is initiated, the disk 
becomes geometrically thin via enormous neutrino emission and starts 
behaving like a Keplerian disk.
Towards the higher end of the above range of $\dot{M}$ in the collapsar~I scenario,
the disk becomes optically thick to the neutrinos in the innermost region. However, still neutrino
cooling is evident in this situation, since after interacting with the hot matter the neutrinos diffuse out
from the system before being accreted into the BH. Photons are trapped and these disks are
optically thick to photons \cite{Kohri}. In the remaining range of $\dot{M}$ of collapsar~I, 
although the disk is optically thin
to neutrinos, they are optically thick to photons and hence geometrically thin too. Thus the inner region 
of the collapsar~I accretion disks mimic the 
celebrated Shakura-Sunyaev model \cite{Shakura} which describes optically thick, geometrically thin 
Keplerian accretion disks. 

The various flow variables (e.g. temperature, flux, density) in the Shakura-Sunyaev model are expressed in terms of 
explicit algebraic formulae.
Novikov \& Thorne \cite{Novikov} worked out the general relativistic version of the Shakura-Sunyaev  
model. Recently, their solution was further reproduced with a more general scaling \cite{Abramowicz2011}.
In the general relativistic Keplerian disk, the flow variables
are expressed in terms of $m=M/M_{\odot}$, $\dot{m}=\dot{M}/\dot{M}_{Edd}$  
where $\dot{M}_{Edd}$ the Eddington accretion rate, $r_*=R/M$ with $R$ being the distance
in the disk from the BH, and the dimensionless Kerr parameter $a_*=a/M$. 

We fix one of the flow variables as reference to establish 
the desired correlation. We also have to choose it in such a way that we can relate it
with observation. 
The analytical forms of these variables have been obtained in the outer, middle and the inner
region of the Keplerian disk. However, except flux ($F$) all have different expressions in 
different regions. 
We thus use $F$ to constrain the mass and spin. The approximate constancy of the neutrino luminosity 
from these neutrino dominated accretion disks further motivated us to choose flux as our reference 
variable.

The flux (here of neutrino) is given by 
\cite{Novikov,Abramowicz2011}
 \begin{equation}
F = [7 \times 10^{26}~{\rm erg~cm^2~s^{-1}}](\dot{m}~m^{-1})r_*^{-3}B ^{-1} C ^{-1/2}Q,
\end{equation}
where 
\begin{equation}
B = 1 + a_*y^{-3},\,\,\
C = 1 - 3y^{-2} + 2a_*y^{-3},\,\,\,y=r_*^{1/2}, 
\end{equation}
and $Q$ is a function of $y$, $M$ and $a_*$ (see supplemental material \cite{sm} for exact expression).
Since $F$ is directly related to the luminosity, which is typically $\sim 10^{52}-
10^{53} \rm ~ergs ~ s^{-1}$ from a neutrino dominated core collapsing disk \cite{smartt},
we use the flux to correlate the mass and spin of SMBHs.

Collapsar~II accretion disks have $0.0001 \lesssim \dot{M}/M_\odot s^{-1} \lesssim 0.01$
\cite{Matsuba, Fujimoto}. Here, the 
core of the progenitor collapses to form a proto-neutron star initially and a mild supernova
explosion is driven. Then a part of the supernova ejecta falls back onto the nascent neutron star which 
subsequently collapses into a BH. 
The temperature in these disks barely attains $10^{10} K$ in the 
inner region and hence these disks are not efficiently cooled by neutrinos. In fact these disks are 
chiefly advection dominated \cite{chen} and hence they can be categorized under general advective 
sub-Keplerian disks.
In this situation, we do not have a simple analytical expression for the various flow variables as in the 
Keplerian disk. Note that self-similar analytical description of sub-Keplerian/advection dominated accretion
flows is applicable only for radiatively inefficient very much sub-Eddington accretion flows \cite{ny94,gm09}. 
Moreover, the neutrino luminosity in the collapsar~II disks is $2-3$ orders of magnitude less than that 
of collapsar~I disks. Consequently, the flux emitted from these disks will also be lower by similar orders 
of magnitude. In order to model such disks, we need to solve the mass transfer, the radial and 
azimuthal momenta balance and the energy equations 
for a sub-Keplerian flow self-consistently and thereby obtain a relation between the mass and spin of
the underlying BHs. The basic equations are given in \cite{gm03,rm10} in detail. 
However, unlike those work, the present work assumes
that the small amount of heat ($\sim 1\%$) escaping from the 
system is carried away by the neutrinos.


\paragraph*{Solution procedure and results$-$}

With all the above ideas in mind, let us explore the said correlation.
Earlier authors \cite{MW99} introduced the parameter $j_{16} \equiv j/\left(10^{16} \rm cm^{2} \rm s^{-1}\right)$, where 
$j$ is the specific angular momentum of the progenitor in CGS unit, and showed that if $j_{16} \lesssim 3$, 
the material falls back into the BH spherically and no disk develops. On the other hand, when
$j_{16}\gtrsim 20$, the centrifugal 
force halts the infalling matter outside $R=1000$km where the neutrino losses are negligible. 
However, for
$3 \lesssim j_{16} \lesssim 20$, an accretion disk forms where neutrino cooling is efficient
\cite{MW99}. The disk then behaves like a Keplerian disk where
the Keplerian radius $R_{Kep}$ is related to $j$ by \cite{MacFadyen}
\begin{equation}
R_{Kep} = j^2 / M.
\end{equation}

We know that in general relativity a stable Keplerian orbit cannot be formed inside a certain radius called  
the innermost stable circular orbit ($r_{ISCO}$). 
For a given mass of the BH, $r_{ISCO}$ is maximum for a non-rotating (Schwarzschild) BH which is $6M$. Thus, if we assume 
that the Keplerian disk extends upto ISCO where $j_{16} \sim 3$ and the BH is 
of Schwarzschild type, then we can have an estimate of the minimum mass of the BH, given by
\begin{equation}
M = j / \sqrt{6},
\end{equation}
which comes out to be $\sim 2.75 M_{\odot}$ after putting various units appropriately. This estimate
gives us an idea of the lower mass limit of BHs, before exploring the mass-spin relation in detail.

To obtain a luminosity $\sim 10^{52}-10^{53} \rm ~ergs ~ s^{-1}$ from 
the collapsar~I disk, the flux is 
accordingly constrained. For example, in order to obtain such an observed luminosity, the neutrino flux 
profile
has to be constrained so that $F\sim 10^{36}-10^{37} \rm ~erg ~ cm^{-2} ~s^{-1}$ at $R \sim 10 M$. 
For the purpose of present computations, we have to fix a value of luminosity from the 
given range and therefore also fix a value of flux at $R \sim 10M$. The value of flux with which we work
is $F\sim 10^{37} \rm ~erg ~ cm^{-2} ~s^{-1}$. 
We fix this $F$ at $R \sim 10M$, vary $\dot{M}$
in the range $10-0.01 M_{\odot} s^{-1}$ (for highly massive to relatively less massive
progenitors), $a_*$ in $1-0$, and finally using these in equation (2)
we determine the mass of the BH. The choice of $R \sim 10M$ assures that at that radius the temperature 
suitable for neutrino emission is achieved in all the cases of $a_*$ considered here, guaranteeing the 
existence of a Keplerian disk always. Hence, repeating the same calculations at other $R$ (fixing
$F$ accordingly), as long as the neutrino cooling has been switched on completely, 
does not alter the main result.

Figure 1 shows the variation of $M$ with $a_*$ using $\dot{M}$
as the parameter. 
For a fixed $\dot{M}$, the mass increases
with the increasing spin of the BH. Thus for a particular $\dot{M}$, the minimum mass of the 
BH arises for $a_*=0$ (Schwarzschild BH) and the maximum mass arises for 
$a_*=1$ (maximally spinning Kerr BH). These minimum and maximum masses get 
enhanced as $\dot{M}$  increases. According to our calculations, the maximum possible
mass of a SMBH is $\sim 85 M_{\odot}$, which is achieved for $a_*=1$ when $\dot{M}=10M_{\odot} \rm s^{-1}$.
It has been already predicted based on the metallicity of the environment 
that the maximum possible mass of SMBHs could vary from $10M_{\odot}$ (for 
super-solar metallicity) to $80M_{\odot}$ (extremely low metallicity)
\cite{Belczynski}.
Thus our theory not only explains the above inferred mass but also predicts the spin of such massive 
SMBHs. Moreover, we can further predict that collapsar~I disks will harbor these BHs,
since they correspond to $\dot{M} \sim 10M_{\odot} \rm s^{-1}$. Hence these BHs should have 
the progenitor mass 
greater than $40 M_\odot$, which is the characteristic of collapsar~I.  

Figure 2 depicts $M$ as a function of $\dot{M}$ using $a_*$ as a parameter. We find that $M$ also increases
with $\dot{M}$ for a fixed $a_*$. This is quite expected because a greater $\dot{M}$ corresponds to a 
greater supply of
mass which finally feeds the BH.

The above discussion is valid only for a Keplerian disk when high neutrino luminosity is possible.
In the collapsar~II regime, which corresponds to the sub-Keplerian, low luminous flows, we do not
enjoy the luxury of analytical computations, rather we have to treat the problem entirely numerically.
Nevertheless, we again fix the flux at $R \sim 10 M$ in such a way that the luminosity obtained 
from the disk, now, is $\sim 10^{48}-10^{49}$ erg $\rm s^{-1}$.
This is expected because the neutrino luminosity from these disks are sub-dominant \cite{MW99}.
For a fixed $\dot{M}$, we vary $a_*$ from 
$0$ to $1$
and for each $a_*$ we vary $M$ unless the required flux is achieved. In this way we 
calculate the variation of $M$ with $a_*$ for $0.001\lesssim\dot{M}/M_{\odot} s^{-1}\lesssim0.003$, which 
has been depicted
vividly in Fig. 3. 
It is worth mentioning that if $\dot{M}$ decreases below $0.0007 M_{\odot} s^{-1}$, even the 
maximally spinning Kerr BHs will have mass below $2 M_\odot$. This implies that BHs with lesser spin 
will have still smaller mass. However, SMBHs of mass below $2 M_\odot$ is not known \cite{Feryal}. 
Thus our theory predicts 
that $\dot{M} < 0.0007 M_{\odot} s^{-1}$ cannot be prevalent in nature in the 
collapsar scenario as this leads to absurd masses for SMBHs.

\begin{table}[h]
\vskip0.2cm
{\centerline{\large Table 1}}
{\centerline{ Mass and spin of known BHs (referred from existing literature)}}
{\centerline{}}
\begin{center}
\begin{tabular}{|c|c|c|}

\hline
$\rm BH ~ Candidate$ & $a_*$ &  $M(M_{\odot})$\\
\hline

 $\rm A0620-00$ & $0.12 \pm 0.19$ \cite{Narayan12} & $6.61 \pm 0.25$ \cite{Cantrell10,Feryal}\\
\hline
$\rm XTE ~J1550-564$ & $0.34 \pm 0.24$ \cite{Narayan12} & $9.10 \pm 0.61$ \cite{Feryal,Orosz02,Orosz03} \\
 & $0.7\pm 0.01$ \cite{bmqpo} & \\
\hline
$\rm GRO ~J1655-40$ & $0.7 \pm 0.1$ \cite{Narayan12} & $6.30 \pm 0.27$ \cite{Feryal,Greene01}\\ 
 & $0.75\pm0.01$ \cite{bmqpo} & \\
\hline
$\rm GRS ~1915+105$ & $0.975 \pm 0.025$ \cite{Narayan12} & $14.0 \pm 4.4$\cite{Greiner01} \\ 
 & $0.68\pm 0.08$ \cite{bmqpo} & \\
\hline
$\rm 4U ~ 1543-47$ & $0.8 \pm 0.1$ \cite{Narayan12} & $9.4 \pm 1.0$ \cite{Feryal,Orosz03}\\ 
\hline
$\rm H ~ 1743-322$ & $0.74 $ \cite{bmqpo} & $11.3$ \cite{bmqpo}\\ \hline

\end{tabular}

\end{center}
\end{table}

In Table I, we present the mass and spin predicted/inferred from observed data
for some of the SMBHs, showing
mass does not increase always with the increasing spin of the BH. This argues for the importance 
of a third parameter, determining the correlation, to relate the mass with spin. According to our
theory, this appears to be $\dot{M}$.
Indeed, Figure 1 shows the strong influence of $\dot{M}$ on $M$ and $a_*$. For example, if GRS~1915+105
has $a_*\sim 0.975$ and $M \sim 14 M_{\odot}$, we can predict from Figure 1 that
during its formation $\dot{M}$ in the collapsar disk was $\sim 0.3-0.5 M_{\odot}s\rm ^{-1}$. For 
GRO~J1655-40,
$a_*$ is lower with lower $M$ compared to that of GRS~1915+105, when $\dot{M}$ is
also lower (slightly less than $0.1 M_{\odot}s\rm ^{-1}$),
as also obtained from Figure 1. However, XTE~J1550-564 has higher $M$ but lower $a_*$ compared to
that of GRO~J1655-40. Hence apparently for GRO~J1655-40 and XTE~J1550-564, the 
spin does not increase with increasing mass. This could only be explained by arguing that $\dot{M}$ at 
the formation stage of XTE~J1550-564
was higher compared to GRO~J1655-40. Indeed, this is confirmed from Fig. 1 when we find that 
$\dot{M}$ for GRO~J1655-40 was $\sim 0.1 M_{\odot}\rm s^{-1}$, whereas for
XTE~J1550-564, it was $\sim 0.3-0.5 M_{\odot}s^{-1}$ during their formation.
On second thought, based on Fig. 3, GRO~J1655-40 can have $0.002<\dot{M}/M_{\odot}\rm s^{-1} 
\lesssim 0.003$ if it had a sub-Keplerian disk at the time of formation. 


Thus, if we know the mass and spin 
of BHs by independent methods, we can predict $\dot{M}$ at their formation stage from Figures 1,
2 and 3, which in turn gives us the information about the hydrodynamics of the underlying 
collapsar accretion disk. This is further related to the observed supernova.

\begin{figure*}
\begin{center}
\includegraphics[angle=0,width=15.5cm]{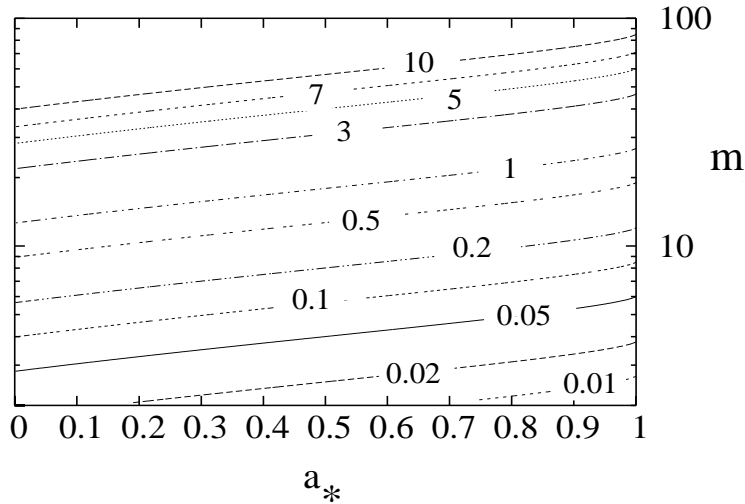}
\caption{ 
Mass as a function of the BH's spin for a collapsar~I disk
using $\dot{M}$ in units of $M_\odot \rm s^{-1}$ as the parameter, labelled in each contour.
  }
\label{eos1}
\end{center}
\end{figure*}

\begin{figure*}
\begin{center}
\includegraphics[angle=0,width=15.5cm]{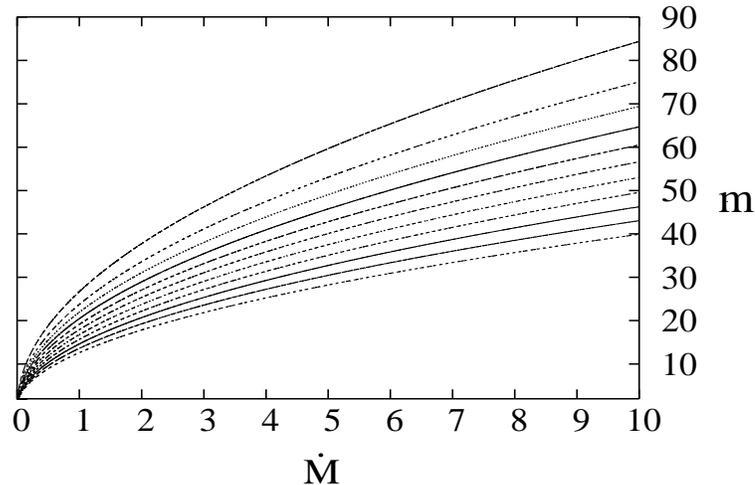}
\caption{ 
Mass as a function of $\dot{M}$ in units of $M_\odot \rm s^{-1}$ for collapsar~I 
disk using the BH's spin as the parameter. From the bottom to top, the contours correspond to
$a_*=0,0.1,0.2,0.3,0.4,0.5,0.6,0.7,0.8,0.9,0.998$.
  }
\label{eos1}
\end{center}
\end{figure*}

\begin{figure*}
\begin{center}
\includegraphics[angle=0,width=11.5cm]{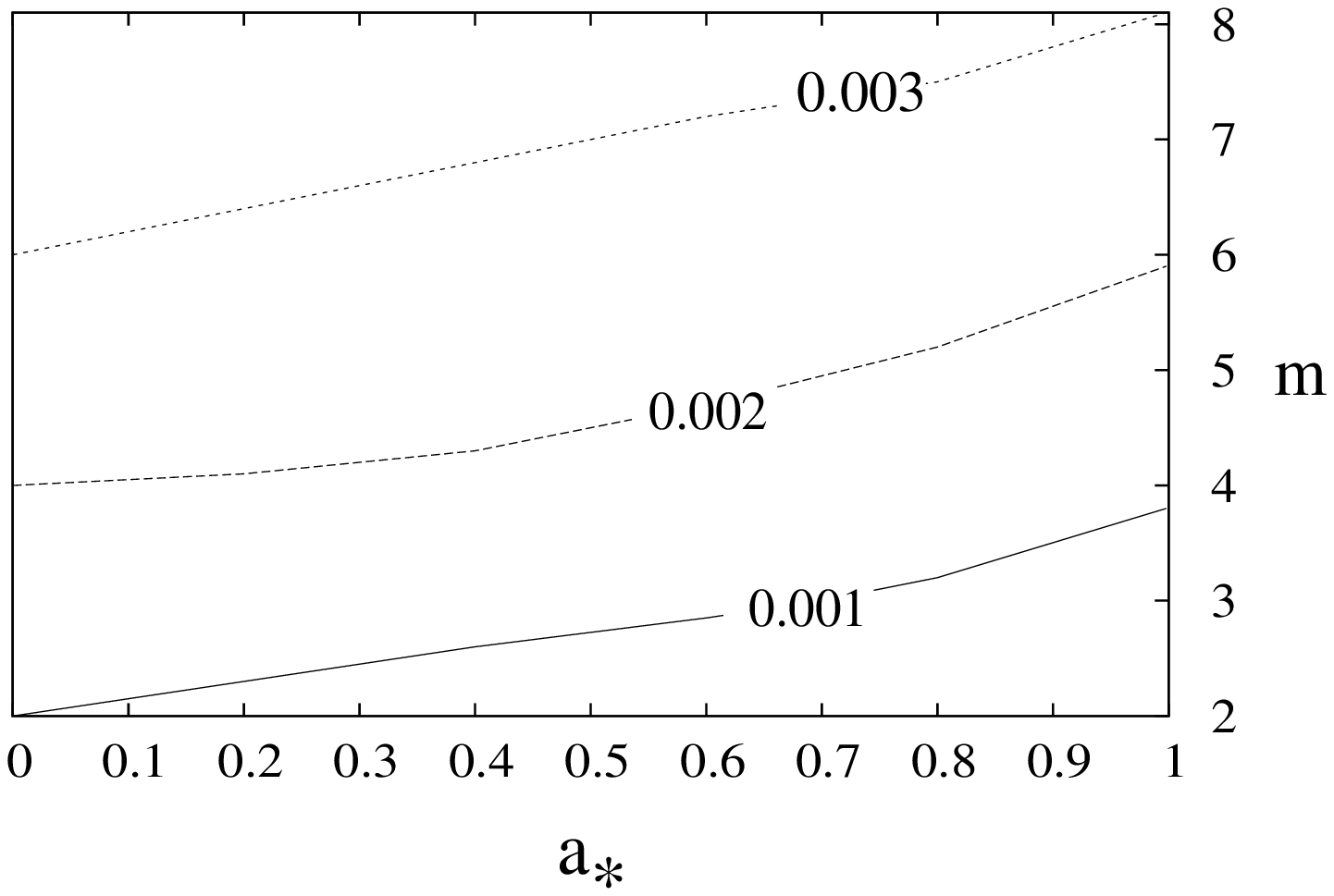}
\caption{ 
Mass as a function of the BH's spin for a collapsar~II disk
using $\dot{M}$ in units of $M_\odot \rm s^{-1}$ as the parameter, labelled in each contour.
  }
\label{eos1}
\end{center}
\end{figure*}

\paragraph*{Conclusions.$-$}
We have found that for a fixed $\dot{M}$, the mass of BHs increases with the increasing spin.
However, from observations we know that there are several BHs having higher spin but lesser
mass with respect to each other. In order to reconcile with those results, we argue that a BH with a higher 
spin can 
have a similar or lesser mass compared to another BH, if the $\dot{M}$ at its formation stage was low 
compared to other 
BH with a lower spin but similar or higher mass. We predict that the maximum
mass of SMBHs can be as high as $85 M_{\odot}$, which tallies with that inferred based on
observed data. Our theory establishes that these BHs are maximally
spinning. Moreover, if the mass and spin of SMBHs are known from observations, we can predict the 
collapsar scenario under which they were formed.
We also provide a range of possible mass for SMBHs and their respective spins, which could be useful
in constraining 
the models predicting spins of SMBHs with known masses.
The least massive BH is a Schwarzschild BH and the most massive one is the extremal Kerr BH.
Interestingly, our theory predicts that a newly formed BH can be maximally spinning, 
which might indicate the formation of a naked singularity if such a BH's spin exceeds
unity via accretion. However, it was argued that \cite{Kesden}, on further accretion $a_*$ 
cannot increase beyond unity. 
In fact, the maximum spin that a BH can attain 
via accretion is $a_* \simeq 0.998$ \cite{Thorne}, 
naked singularities are unstable and a compact object
with $a_* > 1$ will get spun down to a BH state even if it forms \cite{De Felice}. 

The idea proposed here could also be applicable to the observed BHs with known $M$, $a_*$ and 
luminosity (and hence $\dot{M}$) following their evolution after birth,
if they have a donor.
However, such an evolution (of $M$ and $a_*$) is expected to be much slower
due to much lower $\dot{M}$. 
Although BHs in X-ray binaries may be accreting for long times, accretion will 
not significantly affect the natal spin of the BHs \cite{King,McClintock}. 
It was argued \cite{McClintock} that the spin of 
the BH in 4U~1543-47 is chiefly natal considering its present 
$\dot{M}$ and modest age $(\lesssim 1~Gyr)$. It was further reported \cite{McClintock} that 
$a_* \sim 0.98$ in GRS~1915+105 was chiefly 
natal, because achieving this spin gradually via 
accretion would require almost doubling the mass of the BH. 
Such a huge enhancement in BH mass is very unlikely during the 
evolution of GRS~1915+105 or any BH binary. Because, systems
with initially low or moderate mass companions (i.e., $M_c \lesssim {\rm few}~M_\odot$) 
simply cannot supply the required mass, and systems with high-mass 
companions have too short a lifetime to affect the 
required mass transfer. It was also predicted \cite{Lee}  
that GRO~J1655-40 and 4U~1543-47 will have their natal spin $\sim 0.8$,
matching with observational predictions \cite{McClintock}.
All these imply that even if there is a change of mass and spin
of the BHs via their accretion process, this will only
appear as a small correction to the mass-spin contours presented by
us, leaving the overall mass-spin relation intact.
\\ \\

B.M. acknowledges partial 
support through research Grant No. ISRO/RES/2/367/10-11.

\newpage
\numberwithin{equation}{section}
\section*{\large Supplementary information for the letter}

\section{Flux from a general relativistic Keplerian disk} 

The expression for flux ($F$) is described as \cite{Novi,Abramowicz}
\begin{equation}
F = [7 \times 10^{26}~{\rm erg~cm^2~s^{-1}}](\dot{m}~m^{-1})r_*^{-3}B ^{-1} C ^{-1/2}Q,
\end{equation}
where 
\begin{equation}
B = 1 + a_*y^{-3}, 
\end{equation}
\begin{equation}
C = 1 - 3y^{-2} + 2a_*y^{-3}, 
\end{equation}
\begin{eqnarray}
\nonumber Q = Q_0 \left[y -y_0 -\frac{3}{2}a_*{\rm ln}\left(\frac{y}{y_0}\right) - \frac{3(y_1-a_*)^2}
{y_1(y_1-y_2)
(y_1-y_3)}{\rm ln}\left(\frac{y-y_1}{y_0-y_1}\right)\right] \\- Q_0 \left[\frac{3(y_2-a_*)^2}{y_2(y_2-y_1)
(y_2-y_3)}{\rm ln}\left(\frac{y-y_2}{y_0-y_2}\right)+\frac{3(y_3-a_*)^2}{y_3(y_3-y_1)
(y_3-y_2)}{\rm ln}\left(\frac{y-y_3}{y_0-y_3}\right)\right],
\end{eqnarray}
\begin{equation}
Q_0 = \frac{1 + a_*y^{-3}}{y(1-3y^{-2}+2a_*y^{-3})^{1/2}}, 
\end{equation}
\begin{equation}
y=(R/M)^{1/2}, 
\end{equation}
\begin{equation}
y_1 = 2{\rm cos}[({\rm cos^{-1}}a_* - \pi)/3],
\end{equation}
\begin{equation}
y_2 = 2{\rm cos}[({\rm cos^{-1}}a_* + \pi)/3],
\end{equation}
\begin{equation}
y_3 = -2{\rm cos}[({\rm cos^{-1}}a_*)/3],
\end{equation}
\begin{equation}
y_0=(R_{ISCO}/M)^{1/2}, 
\end{equation}
\begin{equation}
R_{ISCO} = M \left\lbrace 3 + Z_2 \mp [(3-Z_1)(3+Z_1+2Z_2)]^{1/2} \right\rbrace,
\end{equation}
\begin{equation}
Z_1 = 1 + \left(1-a_*^2\right)^{1/3}\left[\left(1+a_*\right)^{1/3} + \left(1-a_*\right)^{1/3}\right] \rm 
\end{equation}
and
\begin{equation}
Z_2 = \left(3 a_*^2 + Z_1^2 \right)^{1/2}.
\end{equation}

The symbols $R$, $M$ $\dot{m}$, $m$, $r_*$ and $a_*$ are  already described in the main paper.

\end{document}